\documentclass[preprint,12pt]{elsarticle}

\usepackage{graphicx}
\usepackage{amssymb}
\usepackage{amsmath}

\usepackage{color}

\journal{Journal of Computational Physics}

\def\cfl{C_\mathrm{CFL}}

\begin{document}

\begin{frontmatter}


\title{Enforcing the Courant-Friedrichs-Lewy Condition in Explicitly Conservative Local Time Stepping Schemes}



\author[fnal,kicp,aac]{Nickolay Y.\ Gnedin}
\author[kicp,aac]{Vadim A.\ Semenov}
\author[kicp,aac,efi]{Andrey V.\ Kravtsov}

\address[fnal]{Particle Astrophysics Center, Fermi National Accelerator Laboratory, Batavia, IL 60510, USA; gnedin@fnal.gov}
\address[kicp]{Kavli Institute for Cosmological Physics, The University of Chicago, Chicago, IL 60637 USA}
\address[aac]{Department of Astronomy \& Astrophysics, The University of Chicago, Chicago, IL 60637 USA}
\address[efi]{Enrico Fermi Institute, The University of Chicago, Chicago, IL 60637 USA}

\begin{abstract}
An optimally efficient explicit numerical scheme for solving fluid dynamics equations, or any other parabolic or hyperbolic system of partial differential equations, should allow local regions to advance in time with their own, locally constrained time steps. However, such a scheme can result in violation of the Courant-Friedrichs-Lewy (CFL) condition, which is manifestly non-local. Although the violations can be considered to be ``weak'' in a certain sense and the corresponding numerical solution may be stable, such calculation \emph{does not guarantee the correct propagation speed for arbitrary waves}. We use an experimental fluid dynamics code that allows cubic ``patches'' of grid cells to step with independent, locally constrained time steps to demonstrate how the CFL condition can be enforced by imposing a condition on the time steps of neighboring patches. We perform several numerical tests that illustrate errors introduced in the numerical solutions by weak CFL condition violations and show how strict enforcement of the CFL condition eliminates these errors. In all our tests the strict enforcement of the CFL condition does \emph{not} impose a significant performance penalty. 
\end{abstract}


\end{frontmatter}


\section{Introduction}
\label{S:intro}

In a seminal paper \cite{cfl28} Richard Courant, Kurt Friedrichs, and Hans Lewy showed that any explicit numerical scheme for solving fluid dynamics equations -- or any other parabolic or hyperbolic system of partial differential equations (PDEs) -- can be \emph{stable and converges} to the correct solution only if it satisfies what we now call the Courant-Friedrichs-Lewy (CFL) condition: \emph{"The full numerical domain of dependence must contain the physical domain of dependence"} \cite{laney98}

The CFL condition implies an upper limit on the local time step of a given resolution element in the explicit numerical scheme:
\begin{equation}
	\Delta t\leq \cfl \frac{\Delta x}{v},
    \label{eq:lns}
\end{equation}
where $v$ is the local maximum wave speed in the resolution element, 
but the inverse is not necessarily true -- the local time step constraint does not guarantee that the CFL condition is satisfied. The CFL condition is stricter, since it is \emph{non-local}: the physical domain of dependence also includes waves that can originate somewhere else in the domain and still reach a given location in time $\Delta t$.

The simplest way to ensure that the CFL condition is satisfied is to enforce it globally using the largest wave speed and the corresponding smallest time step in the entire computational volume. However, when the allowed time steps vary widely across the computational volume, this approach can be extremely inefficient because the vast majority of resolution elements are forced to advance with a much shorter time step than allowed by equation (\ref{eq:lns}). 

In the Adaptive Mesh Refinement (AMR) schemes this inefficiency is partly overcome by allowing for larger cells in regions where high resolution and small time steps are not needed. AMR implementations often use ``graded time-stepping'' \cite{bo84,bc89}, whereby the time step is reduced by a fixed factor $\xi_t$ at each subsequent spatial refinement level, with the spatial cell size decrease by a factor of $\xi_s$ and $\xi_t=\xi_s=2$ being a common choice. The graded time-stepping scheme, however, does \emph{does not} eliminate the inefficiency, because the time step in each resolution element still depends on the most restricting time step anywhere else in the solution. For example, if the time step in the most restricting resolution element is decreased, all time steps in all other resolution elements need to be decreased as well.

A more efficient approach would be to allow individual resolution elements, or small localized groups of resolution elements, to advance with their own time steps set by the \emph{local} constraint (\ref{eq:lns}). Such local time stepping can also make parallelization and load balancing more flexible by reducing dependencies between separate domains advanced in parallel. Performance gains and the ease of achieving them likely depend on the specific computational problem and the subject field. This approach does appear promising in cosmological and galaxy formation simulations, where the fraction of volume that requires high resolution and small steps tends to be very small. 

Local time-stepping (LTS) is a relatively new approach in computational physics, although some early attempts to implement it in numerical schemes date to previous century (c.f.\ \cite{os83,s85,c92,ztrc94,d95,sv00}).
Nevertheless, there already exist several thousands research papers discussing various applications of local time stepping, from multi-rate ODE solvers to full PDE schemes with mesh refinement. While giving a full review of this rapidly developing field is beyond the scope of this paper, several excellent recent reviews provide a comprehensive coverage of the current state of the art \cite{Castro2009,Gander2013,chen2013,blazek2015,GOTTLIEB2016,Cohen2017}.

In this paper we focus on a subset of the LTS schemes that are explicitly conservative (i.e.\ maintain the conservation of physical quantities to the machine precision by appropriately tracking and exchanging fluxes between interfaces). 

The local time step constraint (\ref{eq:lns}) can be derived analytically for many simple numerical schemes as a requirement for (linear) numerical stability. One can therefore imagine a numerical scheme in which each resolution element steps in time with its own local maximally allowed time step (\ref{eq:lns}). Such a scheme does not necessarily satisfy the CFL condition and is not guaranteed to be numerically stable for non-linear solutions. Numerical stability can be assured by enforcing the so-called ``update criterion'' (also sometimes called ``evolve condition'') \cite{lgm07,dkt07,lgm08,d14,cdm15} in addition to the local time step constraint (\ref{eq:lns}). The update criterion forbids resolution elements or local domains to advance beyond the next time moment of any of its neighbors:
\begin{equation}
  t_i^{n+1} \leq \min\left\{t_j^{n+1}\right\},~ j\in N_i,
  \label{eq:uc}
\end{equation}
where $N_i$ is the set of neighbors for the resolution element $i$. However, as we discuss below, the update criterion does not, in general, guarantee that the CFL condition is satisfied, and a more stringent condition that we derive below needs to be enforced to guarantee that the CFL condition is satisfied exactly at all times. 

Our approach is, perhaps, most similar to that of \cite{cs07,sc09}, who introduced a buffer region between parts of the solution stepping with different local time steps. However, we argue in the following section that this is necessary, but not sufficient for the strict enforcement of the CFL condition. The latter also requires the capability to reduce the time step of the slower part of the solution "mid-step" in response to the evolving fast part of the solution.

\section{Enforcing the CFL Condition}
\label{S:caus}

In this paper, we use examples based on an experimental fluid dynamics code that follows evolution of the Euler equations on a uniform grid using a conservative Godunov-type scheme. The grid is split into patches -- cubic groups of cells that share the same time step -- for efficiency and ease of implementation. Neighboring patches exchange fluxes and other relevant information, so that global solution is obtained over the entire  domain. Although we present results for this specific implementation of fluid dynamics solver, we believe the conclusions we draw are general and do not depend on the patch size or even shape. Our conclusions also equally apply to an adaptively refined grid.

\begin{figure}[ht]
\centering%
\includegraphics[width=0.75\hsize]{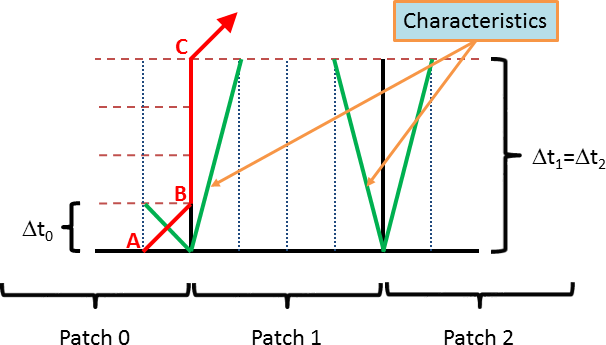}%
\caption{Illustration of artifacts arising when the local time step constraint (\ref{eq:lns}) is satisfied, but the CFL condition is not. Three patches of 4 cells each (cells are shown by the vertical dotted lines)  step in time with unequal time steps, shown with horizontal brown dashed lines: patch 0 makes 4 steps, while patches 1 and 2 make just 1 step. Local time constraints are satisfied in each patch, and so characteristics (green lines) in each patch traverse one cell in one time step; for simplicity, we consider a case with the CFL number of 1. A wave that starts at the moment A reaches the patch boundary at the moment B, but does not propagate into patch 1 until the moment C.
\label{fig:wc}}
\end{figure}

In figure \ref{fig:wc} we illustrate the potential numerical artifacts that can arise if only the local time step constraint is imposed without the strict enforcement of the CFL condition. The sketch shows the ``space-time diagram'' for three grid ``patches,'' along with grid cells (vertical dotted lines). The local time step constraint in patch 0 is 4 times stricter than in the other two patches. Thus, patch 0 makes 4 time steps, while the other two make just one step. The green lines show characteristics of the PDE solution. Here, for illustration, we assume that the scheme uses $\cfl=1$; for $\cfl<1$ a characteristic would traverse a $1/\cfl$ fraction of a cell. In all these cases, characteristics of the PDE propagate the length of one cell or less.  

Problems arise when a wave crosses the boundary between the two patches with different time steps. A wave following the red world line starts at space-time point A, reaches the boundary between patches 0 and 1 at space-time point B, but enters patch 1 only at space-time point C, because patch 1 makes a single time step from the moment $t_A$ to the moment $t_C$. The world line of the wave, therefore, becomes incorrect and incurs a time delay of $t_C-t_B$.

We will call such behavior a ``weak CFL violation,'' to emphasize that the physical causality is not violated because $t_C-t_B>0$ and patch 0 does affect patch 1 in the future. However, due to the time stepping choice, the actual timing of the wave propagation across the patch boundary is incorrect. 

\begin{figure}[t]
\centering%
\includegraphics[width=\hsize]{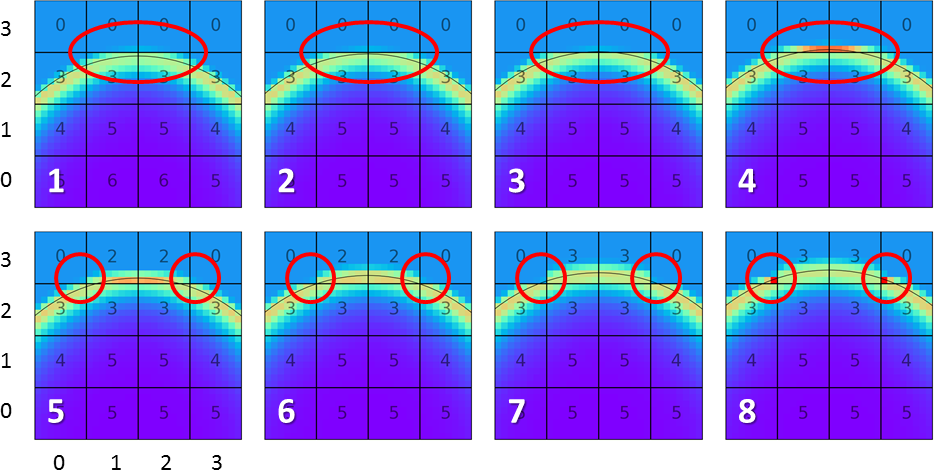}%
\caption{Solution of the 2D point explosion test (see \S\ \ref{sec:tests}) under only the local time step constraint (\ref{eq:lns}). Thin black lines show the analytical solution for the location of the shock. Each of the eight  panels shows a subsequent snapshot in time. Squares are individual patches, with numbers labeling the time step bin. In the top row the artifacts, emphasized by red ellipses, appear at the bottom edges of the patches with lower-left-corner coordinates (1,3) and (2,3), with our chosen coordinate system shown on the outside. In the bottom row these artifacts gradually disappear, but new ones appear in the corners of patches with coordinates (0,3) and (3,3). 
\label{fig:wcsed}}
\end{figure}

The artifacts that can arise from the weak CFL violation are illustrated in figure \ref{fig:wcsed}, which shows a segment of a shock wave resulting from a point explosion in a uniform cold gas. The numerical experiment shown in the figure is evolved in 2D and uses square patches of $8\times8$ cells. To ensure that neighboring patches evolving with different time steps always synchronize at the end of the largest of their time steps, the steps of patches are quantized in fractional powers of two of a single, arbitrary chosen global time step $\Delta t_g$. Note that after $\Delta t_g$ all patches are synchronized. Such synchronization is not generally required, but is useful to produce a synchronized simulation output. 

Integer numbers inside the patches shown in the figure indicate their time step bins. For example, at the first snapshot a patch with lower-left coordinates of (0,3) is in 0th time step bin, i.e. its local time step $\Delta t_{(0,3)} = \Delta t_g$, while a patch with coordinates (3,0) belongs to the 5th time step bin, i.e.\ its local time step is $\Delta t_{(3,0)} = 2^{-5}\Delta t_g = \Delta t_g/32$.

The top four panels show a single global step $\Delta t_g$. As the shock wave reaches the top boundary in the patches with coordinates (1,2) and (2,2), it gets delayed there just like a red line in fig.\ \ref{fig:wc} between points B and C, because zero-bin patches (1,3) and (2,3) advance only in panel 4, which corresponds to the point C in fig.\ \ref{fig:wc}. Thus, the shock wave enters the new patches in panel 4, but the solution remains distorted, because the time delay results in the pile-up of gas in a single row of cells. As the simulation progresses, the pile-up disperses and the numerical solution recovers to the correct one, but now new, similar, artifacts appear at the corners of patches (0,3) and (3,3), for the same reason.

\begin{figure}[th]
\centering%
\includegraphics[width=0.75\hsize]{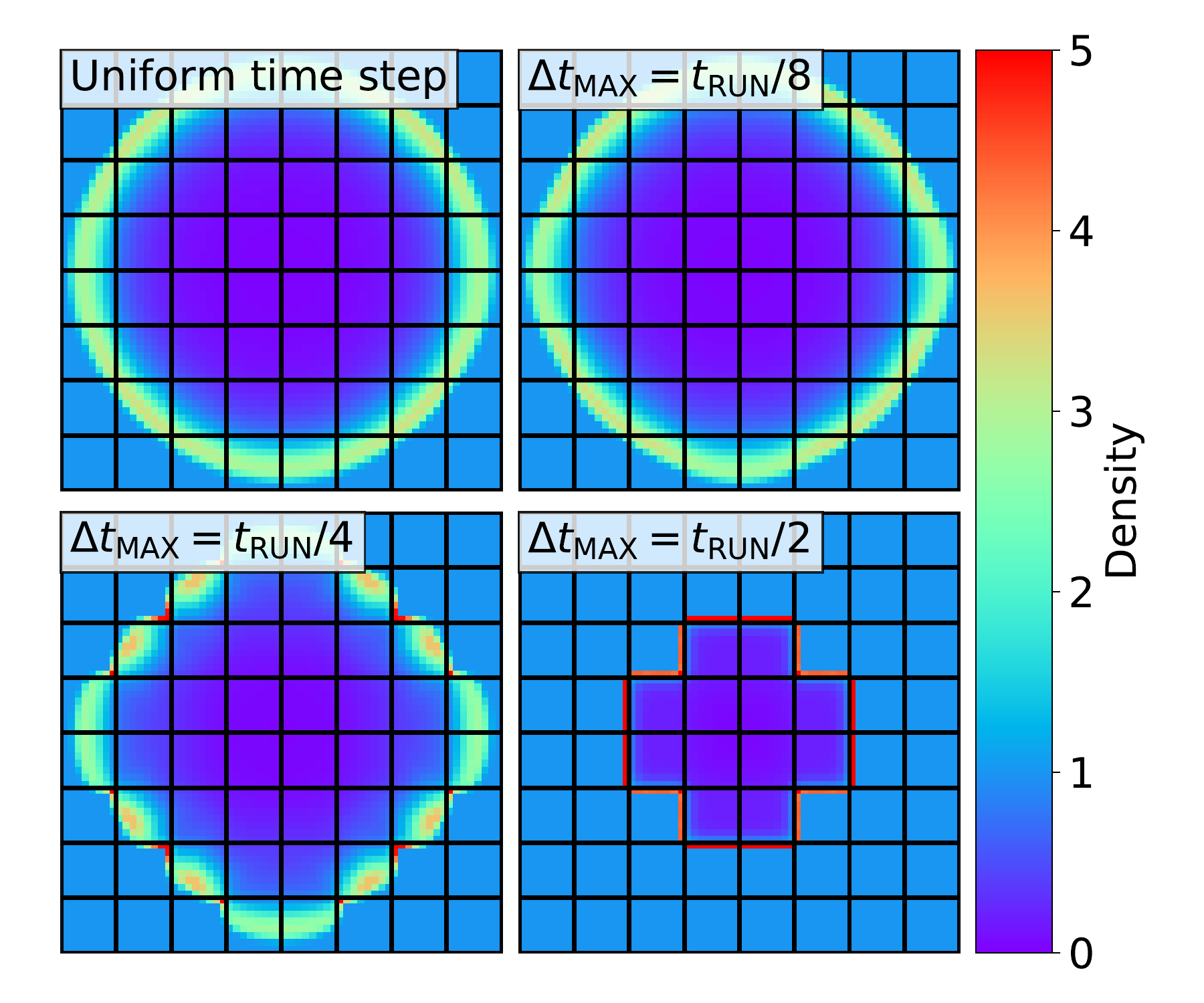}%
\caption{Solutions of the 2D point explosion test (see \S\ \ref{sec:tests}) with different time stepping choices. The top left panel shows the reference solution with globally uniform time steps. The other three panels show LTS solutions with time steps constrained by the local time step constraint (\ref{eq:lns}) and an additional, globally enforced absolute maximum on the value of the local time step. In the three panels these values are 1/8, 1/4, and 1/2 of the total simulation time (chosen as time the shock front takes to reach the edge of the simulation box) respectively, as labeled in the legend. Without additional constraints, artifacts shown in fig.\ \ref{fig:wcsed} can become arbitrary large.
\label{fig:wcsed2}}
\end{figure}

Such artifacts, in principle, can be arbitrary large. For example, consider the point explosion test discussed above, with the initial state at arbitrary cold temperature. If the explosion happens in patch A, all neighboring patches have  arbitrary large local time steps, and, hence, without any additional restriction, they can simply take one time step from the initial moment to the end of the simulation, and the initial shock wave would never leave patch A. In other words, the delay in the wave propagation shown in figure \ref{fig:wc} as the time interval B-C can be arbitrary large. Clearly, such solution is grossly unphysical.

Figure \ref{fig:wcsed2} illustrates this further by showing the reference simulation with the globally uniform time step (the top left panel) and three LTS solutions, in which local time steps are additionally limited by a fraction of the total simulation time. As the time step limit decreases, the LTS solution approaches the reference solution with the globally uniform time step, but for simulations where that limit is too large (like the bottom right panel in fig.\ \ref{fig:wcsed2}), the LTS solution may be grossly wrong. The reason for these large errors is the violation of the CFL condition. It is also worth noticing that the violation of the CFL condition is always in the same direction: waves are delayed (as shown in fig.\ \ref{fig:wc}) but are never sped up (the latter would violate causality). Hence, even small artifacts can accumulate over the course of a long solution into a significant error.

\begin{figure}[th]
\centering%
\centering\includegraphics[width=0.75\hsize]{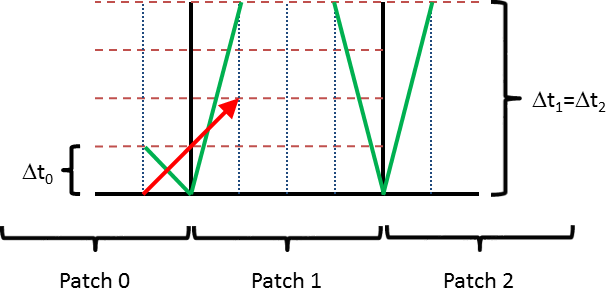}%
\caption{Sketch shown in fig.\ \ref{fig:wc}, but now with equal time steps imposed for neighbors (patch 1) of a resolution element (patch 0) that steps with its maximally allowed time step. Such a restriction ensures correct wave propagation (red line) and, hence, enforces the CFL condition, but only if time steps remain constant.
\label{fig:sc1}}
\end{figure}

In order to enforce the CFL condition and, thus, restore correct propagation for all waves, one can impose an additional requirement when choosing the time step for a given resolution element (c.f.\ patch 1 from figure \ref{fig:sc1}), namely that
\begin{equation}
\begin{minipage}{0.9\textwidth} 
\emph{the time step of any resolution element must satisfy not only its own local time step constraint, but also the local time step constraints of all its neighbors.}
\end{minipage}
\label{eq:ncfl}
\end{equation}
In comparison to figure \ref{fig:wc}, patch 1 now steps with the same time step as patch 0, and the wave shown with the red arrow in both figures now propagates correctly from patch 0 to patch 1. A similar approach was used in \cite{cs07,sc09}, who introduced a buffer between regions with small and large time steps. In the language of fig.\ \ref{fig:sc1}, patch 1 plays a role of such buffer.

\begin{figure}[ht]
\centering%
\includegraphics[width=\hsize]{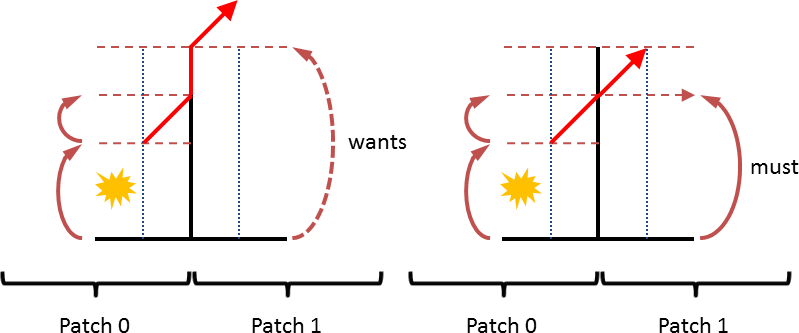}%
\caption{Illustration of violation of the CFL condition when a particular patch (0) experiences an event that reduces its time step (left panel). Ensuring that none of its neighbors step past the end of its next time step (right panel) is a requirement for enforcing the CFL condition.
\label{fig:sc2}}
\end{figure}

However, figure \ref{fig:sc2} shows that the requirement (\ref{eq:ncfl}) is necessary, but not sufficient if the numerical solution contains discontinuities. In this case, patch 0 steps with half the time step of patch 1, but its time step is less than maximally allowed by equation (\ref{eq:lns}); instead the time step of patch 0 is determined by some other neighbor that steps with its own maximally allowed time step. As the solution evolves, patch 0 experiences an event that reduces its own maximally allowed time step, such as a collision of two shocks or a supernova explosion (or any other discontinuity in the solution). In this case the violation of the CFL condition occurs \emph{after} the time step for patch 1 had already been set by the requirement (\ref{eq:ncfl}). 
In order to avoid artifacts shown in the left panel of figure\ \ref{fig:sc2}, the time step of patch 1 has to be reduced ``mid-step,'' as is shown in the right panel of the figure.

Thus, we can formulate a requirement that enforces the CFL condition and thus guarantees correct propagation for all waves:
\begin{equation}
\begin{minipage}{0.9\textwidth} 
\emph{If the local maximally allowed time step for a particular resolution element at any time $t=t_0$ is $\Delta t$, all immediate neighbors of this resolution element must end their time step at or earlier than $t_1=t_0+\Delta t$.}
\end{minipage}
\label{eq:cfl}
\end{equation}
I.e., for a wave to propagate from one region to a neighboring region at a correct time, the regions must be synchronized in time after they complete their maximally allowed steps. This condition ensures that any neighboring regions step frequently enough to avoid delays in wave propagation through the interface between these regions.

\begin{figure}[t]
\centering%
\includegraphics[width=0.75\hsize]{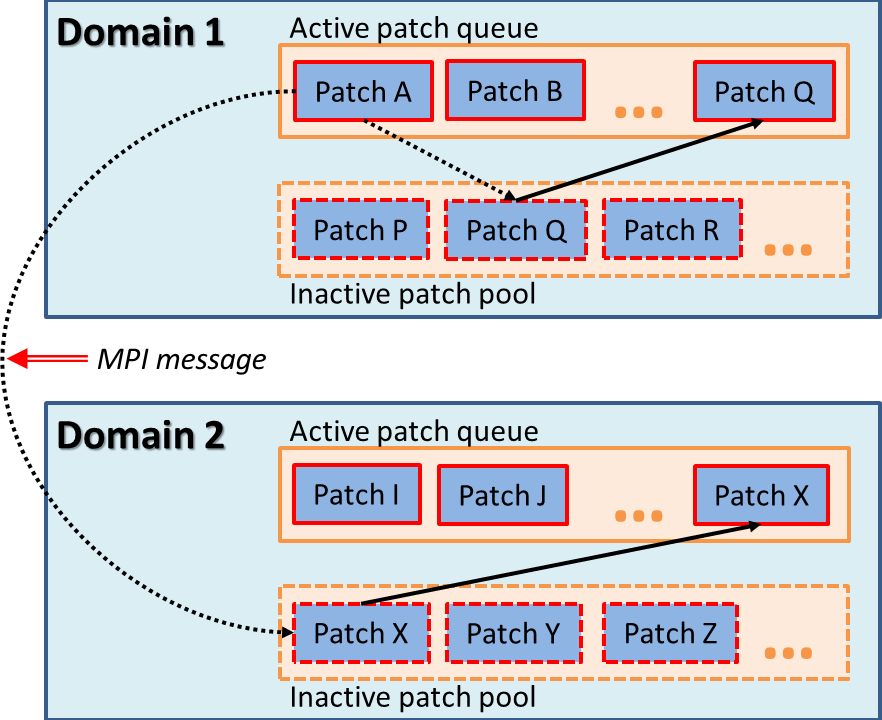}%
\caption{Algorithmic enforcement of the CFL condition in our experimental code. Dotted lines indicate ``pull back'' messages, while solid lines show a simple patch move (a pointer assignment) from an inactive pool into the active queue.}
\label{fig:algo}
\end{figure}

Our experimental code implements one possible way to enforce the CFL condition exactly in practice, as is shown in figure \ref{fig:algo}. The whole computational volume is divided into separate non-overlapping domains that communicate using Message Passing Interface (MPI) library.  Each domain maintains a queue of active patches, whose current time steps end exactly at the time $t$ when the domain ends its step, and a pool of inactive patches whose current time steps end at later times. If an active patch A steps with its own maximally allowed time step, it must ensure that all its neighbors are active too.\footnote{Note however, that if patch A has a time step shorter than required by equation (\ref{eq:lns}), some of its neighbors could be inactive.} If neighbors of patch A include some inactive patches Q and X, these must be activated to guarantee that the CFL condition is satisfied. 

Thus, patch A sends ``pull back'' messages to patches Q and X. Patch Q, which happens to be located in the same domain and within the same MPI rank, is simply moved into the active patch queue of domain 1 via a pointer copy and an update of the end time for patch Q. To activate patch X, an MPI message is sent to patch X, which, in turn, is moved into the active queue of domain 2. 

\begin{figure}[t]
\centering%
\includegraphics[width=0.75\hsize]{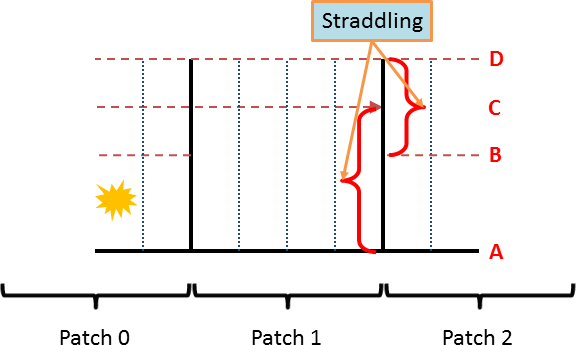}%
\caption{Example of ``straddling'' of two patches. Both starting and finishing times for patch 1 (A$\rightarrow$C) are earlier than respective times for neighboring patch 2 (B$\rightarrow$D).
\label{fig:strad}}
\end{figure}

Such an exchange of messages is, in principle, sufficient to enforce the CFL condition exactly. However, this results in what we call ``straddling,'' as can be illustrated for the previously discussed example shown in figure\ \ref{fig:sc2}. Here patch 1 has a neighbor patch 2 advancing with the twice smaller time step, just like patch 0 before the sudden change in the solution that forces time step change, as shown in figure \ref{fig:strad}. As patch 0 sends a ``pull back'' message to patch 1 and patch 1 reduces its time step to finish at $t_C$, its end time becomes ``straddled'' between the starting ($t_B$) and ending ($t_D$) times of patch 2.

\begin{figure}[ht]
\centering%
\includegraphics[width=0.75\hsize]{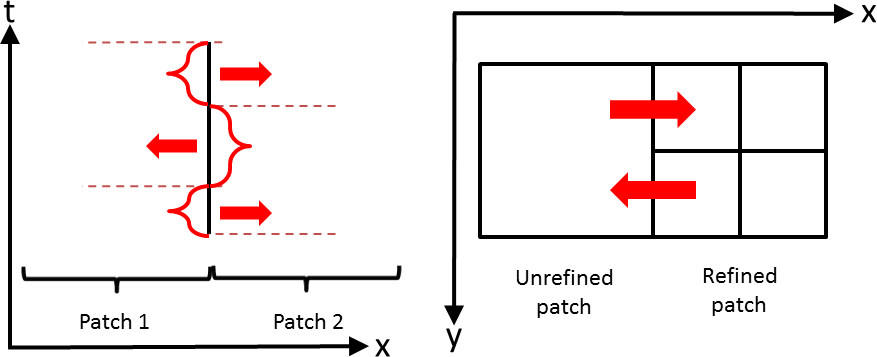}%
\caption{Potential complications of allowing straddling for conservative schemes. Left panel shows an example when fluxes of conserved variables need to be passed from patch 1 to patch 2 and from patch 2 to patch 1 interchangeably. The right panel shows the case with mesh refinement when a single refined interface needs to pass fluxes in two opposite directions. Such schemes can be implemented, but they require complex bookkeeping.
\label{fig:strad2}}
\end{figure}

For conservative numerical schemes straddling complicates proper exchange of fluxes between neighboring patches, as is illustrated in figure \ref{fig:strad2}. However, this is a complication, not a limitation, and with sufficient bookkeeping the issue can be resolved and the CFL condition enforced in a numerical code that allows straddling \cite{lgm07,dkt07}.

In the experimental code that we use in this paper, we adopt a simpler but less efficient approach. In the situation depicted in figure\ \ref{fig:strad}, instead of complex bookkeeping, we force patch 2 to decrease its time step too, so that it ends at the end time of patch 1 step, $t_C$. The actual forcing is implemented by another type of message, ``step drop,'' that patch 1 sends to patch 2.

On a distributed system, this approach encounters a challenge of detecting when all ``drop'' messages are delivered. In the example from figure\ \ref{fig:strad}, when patch 2 drops its time step to $t_C$, it must also send ``drop'' messages to all its other neighbors, because a new straddling condition may arise with one of its other neighbors. Such a wave of ``drops'' may propagate over a substantial part of the whole computational volume. The wave of messages is guaranteed to converge, because time steps are quantizes into powers-of-two of the global time step, and local time steps can thus be represented by integer numbers. In our code we label the smallest time step in the whole domain as 1. Hence, propagation of a wave of drop messages from one patch to another decreases integer labels of their steps by a discrete amount, but each label never decreases below 1. Moreover, it does not have to reach 1 if local conditions allow it to enforce the CFL condition with larger time steps. Clearly, such a wave can propagate only over a finite number of step decreases. However, detecting algorithmically when the wave stops is a non-trivial task. 

We have resolved this challenge by allowing separate domains to synchronize with each other. After domains have sent out all of their "drop" messages, they inform their neighbors with special ``sync'' messages. Domains are not allowed to proceed until they receive ``sync'' messages from all their neighbors. Experimentation showed that one cycle of such ``sync'' messages is not enough because the ``drop'' wave may propagate beyond the nearest neighbors, bounce off and return to the original domain later. Nevertheless, we find that two cycles are sufficient. We do not have a formal proof of this statement, but all our tests and production runs on a realistic physical model described in the following section perform successfully with two rounds of ``sync'' messages. Such synchronization imposes only a minimal cost, since it takes place in parallel with physics calculations on the patches, allowing almost perfect overlap of computation and communication, and synchronization only affects domains, not patches (i.e. the number of message exchanges is small, twice the number of neighbors per domain).

Since our goal in this paper is to illustrate the enforcement of the CFL condition with local time steps, and not to discuss a specific software implementation that addresses the numerical artifacts related to CFL violation, our experimental code is sufficient for this purpose. Undoubtedly, it can be optimized further and inefficiencies, including synchronization between domains, built-in for simplification, can be reduced or eliminated altogether with additional effort. In order to estimate the costs incurred and gains achieved in our scheme, we develop in \ref{sec:app} a simple quantitative model of our LTS scheme and compare it with the canonical globally uniform time stepping scheme.

\section{Performance Tests}
\label{sec:tests}

LTS schemes can result in a large gain in performance when the range of dynamical time scales of the solution within the computational volume is wide and the fraction of cells requiring short time steps is small. As an illustration, we consider a numerical solution of a point explosion in a uniform medium, also known as the Sedov-Taylor problem \citep{neumann1941,sedov1946,taylor1950}. 

The solution in this problem exhibits a large range of gas temperatures and local maximally allowed time steps. In addition, there exists an analytical solution, so numerical errors can be quantified exactly.
To evolve the solution numerically, we use our experimental patch-based code with the gas dynamics solver adopted from the publicly available cosmological code RAMSES \cite{t02}. The details of the solver are presented in \S\ {ref:axcode}

\begin{figure}[ht]
\centering%
\includegraphics[width=0.5\hsize]{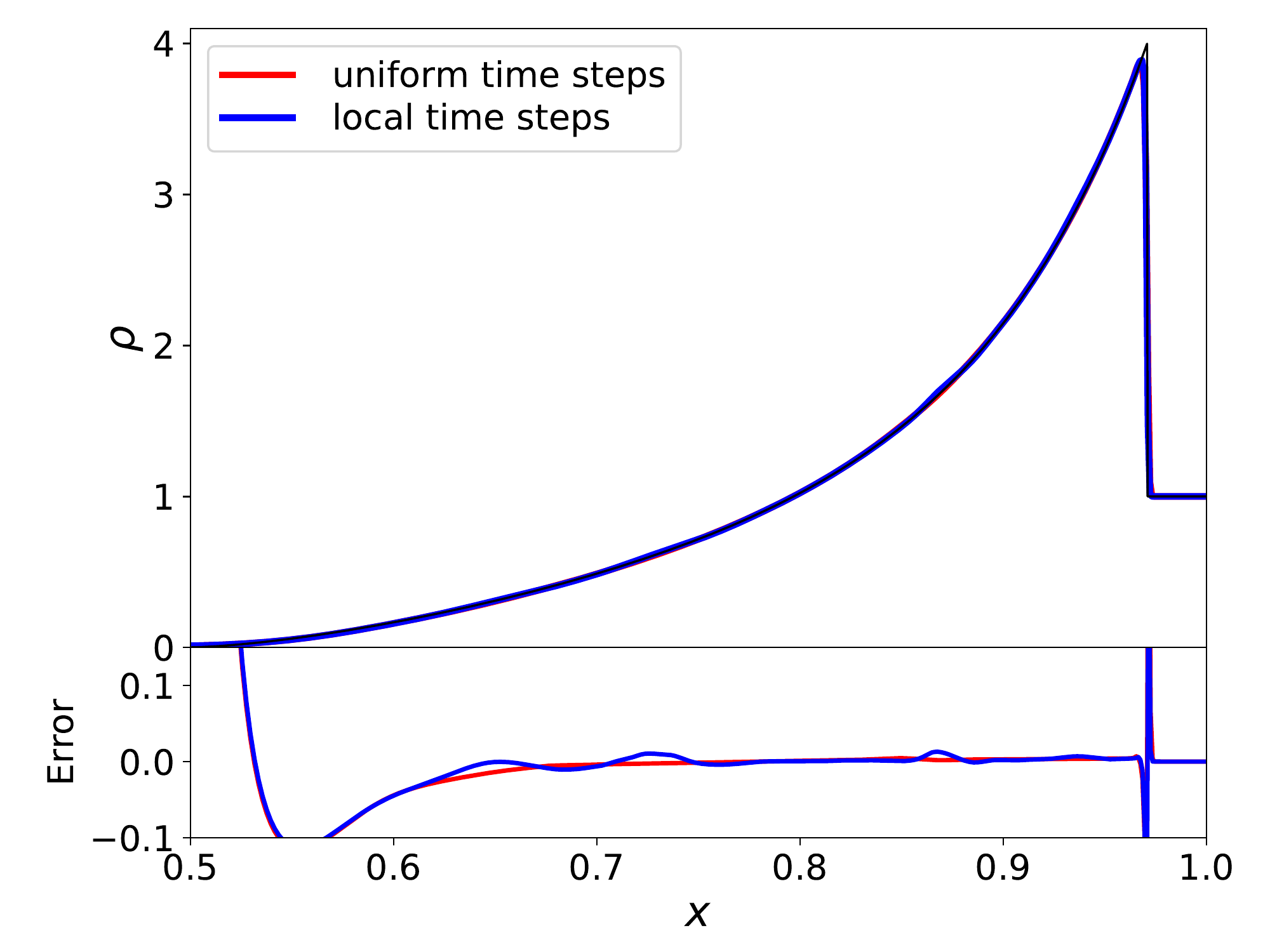}%
\includegraphics[width=0.5\hsize]{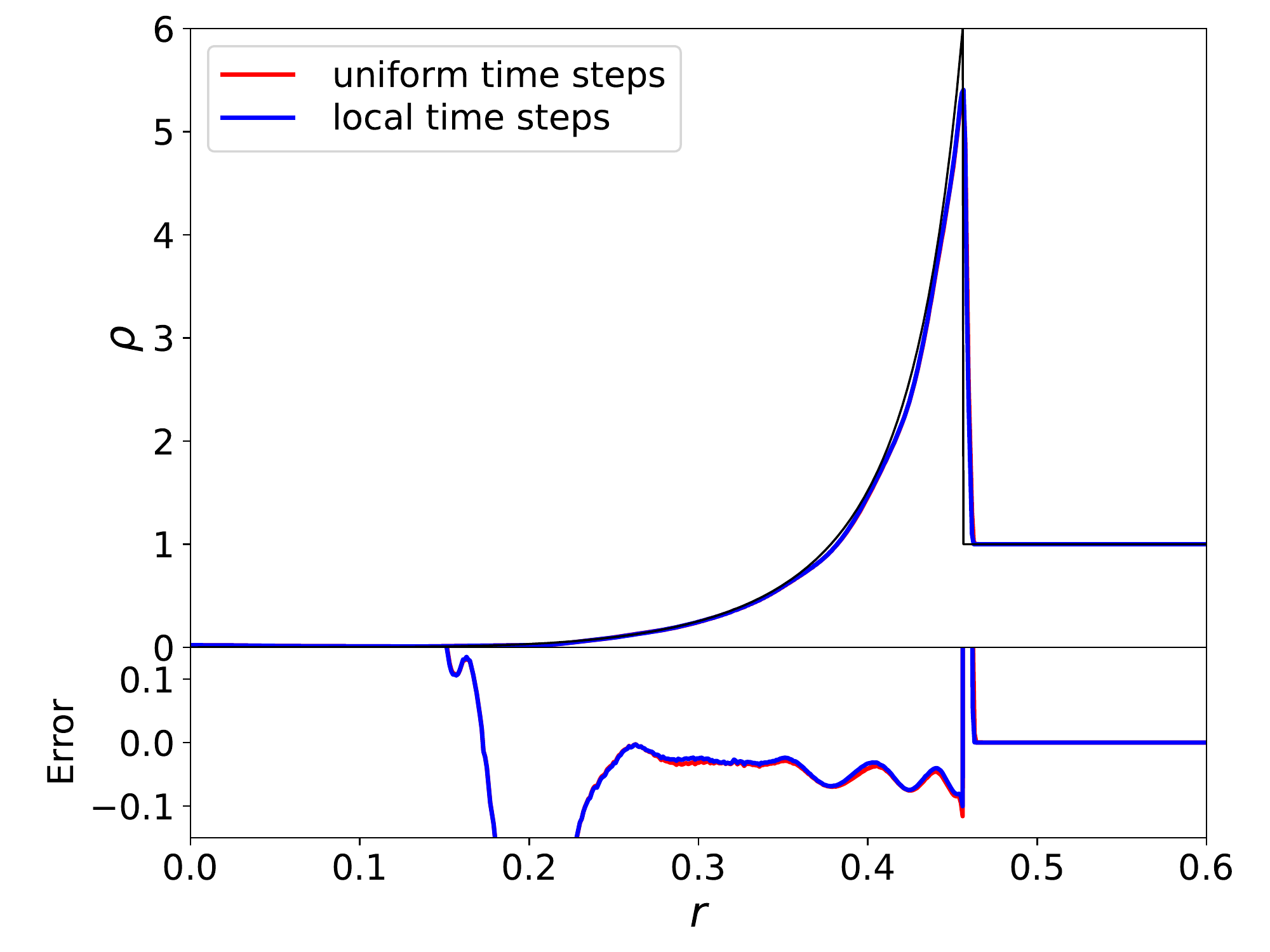}%
\caption{Linear 1D (left) and azimuthally averaged 2D (right) profiles of gas density in the point explosion test with uniform time stepping (red lines) and with local time steps with the enforced CFL condition (blue lines). Black solid lines show analytical solutions, and lower panels show errors of the numerical solution. Two numerical solutions are virtually indistinguishable.}
\label{fig:sed}
\end{figure}

In figure \ref{fig:sed} we show gas density profiles in the one- and two-dimensional numerical solutions for the point explosion test. Both tests were performed with 1024 cells along each dimension, divided between 128 patches of 8 cells each (i.e.\ in 1D we used $128$ patches of $8$ cells, and in a 2D case we used $128\times128$ patches of $8\times8$ cells each).

In each case we compare two solutions, obtained with globally uniform and local time stepping. The figure shows that the two solutions are virtually indistinguishable. We have also verified that the numerical convergence rates of both solutions are also essentially identical.

\begin{figure}[ht]
\centering%
\includegraphics[width=0.75\hsize]{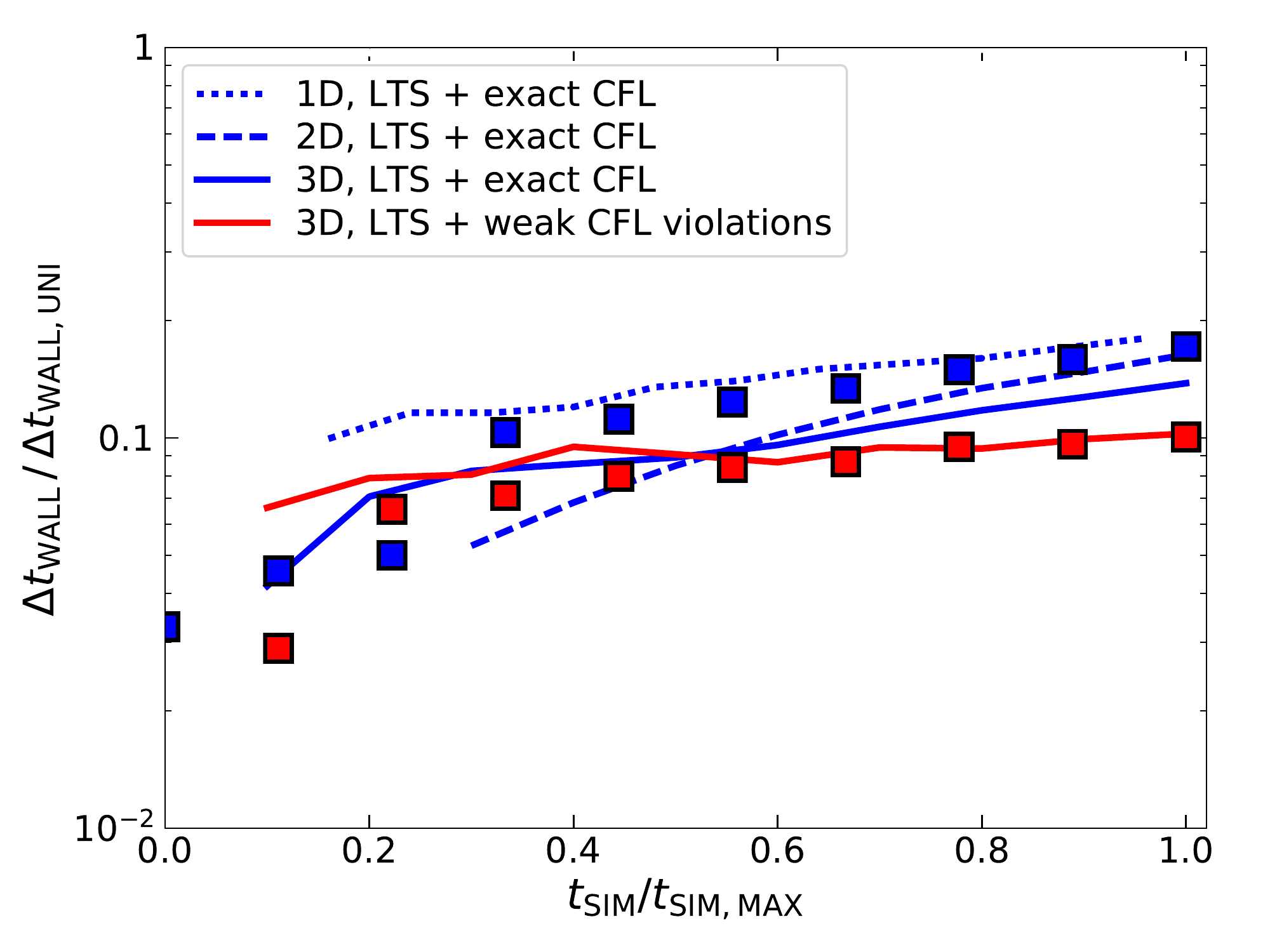}%
\caption{The ratio of the wall-clock time per one global time step for 1D, 2D, and 3D point explosion tests for runs with local time steps and runs with uniform time stepping vs the simulation progress. Local time stepping runs are $\sim 6-10$ times faster in all cases. Color squares show predictions of the execution model (\ref{eq:rltsflux}) for the 3D case with time step bin fractions $f_j$ and execution times $N_B T_C$ and $N_VT_W$ measured directly from the simulations.}
\label{fig:tsed}
\end{figure}

Figure \ref{fig:tsed} shows the ratio of the wall-clock time per each global time step in the local and globally uniform time stepping runs of the same dimension. Runs with local time stepping are $\sim 6-10$ times faster than the runs with uniform time stepping for solutions of all dimensions, in good agreement with the predictions of our execution model from \ref{sec:app} (equation \ref{eq:rltsflux} - the model is somewhat above the actual data for the exact CFL case, but, after all, it is formally an upper limit).

Even more remarkably, the speed gain seems to largely persist until the end of each test, despite the fact that we continue our tests until the shock wave reaches the boundary. Naively, one would expect the gain in performance to be large initially, when the cold unshocked gas has long local time steps, while the time steps in hot shocked gas are small. However, as the shocked region expands to encompass most of the computational volume, the benefit of the local time stepping could be expected to diminish. As fig.\ \ref{fig:tsed} illustrates, this is not the case: there is a sufficient range in the local time step conditions in the shocked gas to result in a substantial performance gain even when most of the computational volume is shocked.

A solution that allows weak CFL violations is even faster, but may exhibit artifacts shown in fig.\ \ref{fig:wcsed} and \ref{fig:wcsed2}. In order to eliminate these artifacts, we limit the maximum allowed time step for the weakly CFL violating solution to $1/200$ of the simulation duration - the weakest such restriction that keeps artifacts visually insignificant. Such a restriction, however, eliminates (in this particular test) most of the speed gain that may be achieved by violating the CFL condition.

For a more complex and realistic test, we use a simulation of decaying isothermal turbulence, a common numerical problem in modeling interstellar medium and star-forming regions in astrophysics. Initial conditions for this test is a snapshot from the comparison project of astrophysical fluid dynamics codes reported by \cite{knc11}. The data represent the state of the gas in a cubic volume covered with the uniform $256^3$ grid. In the simulation of \cite{knc11}, turbulence was forcefully driven for some time, and then allowed to decay after driving was switched off. The snapshot we use was taken at $t=0.02$ after the end of driving, where time is measured in units of the sound crossing time in the box. We continued the simulation from $t=0.02$ to $t=0.1$, when the turbulence decayed appreciably but not completely, making it a reasonably realistic test.

\begin{figure}[ht]
\centering%
\includegraphics[width=0.75\hsize]{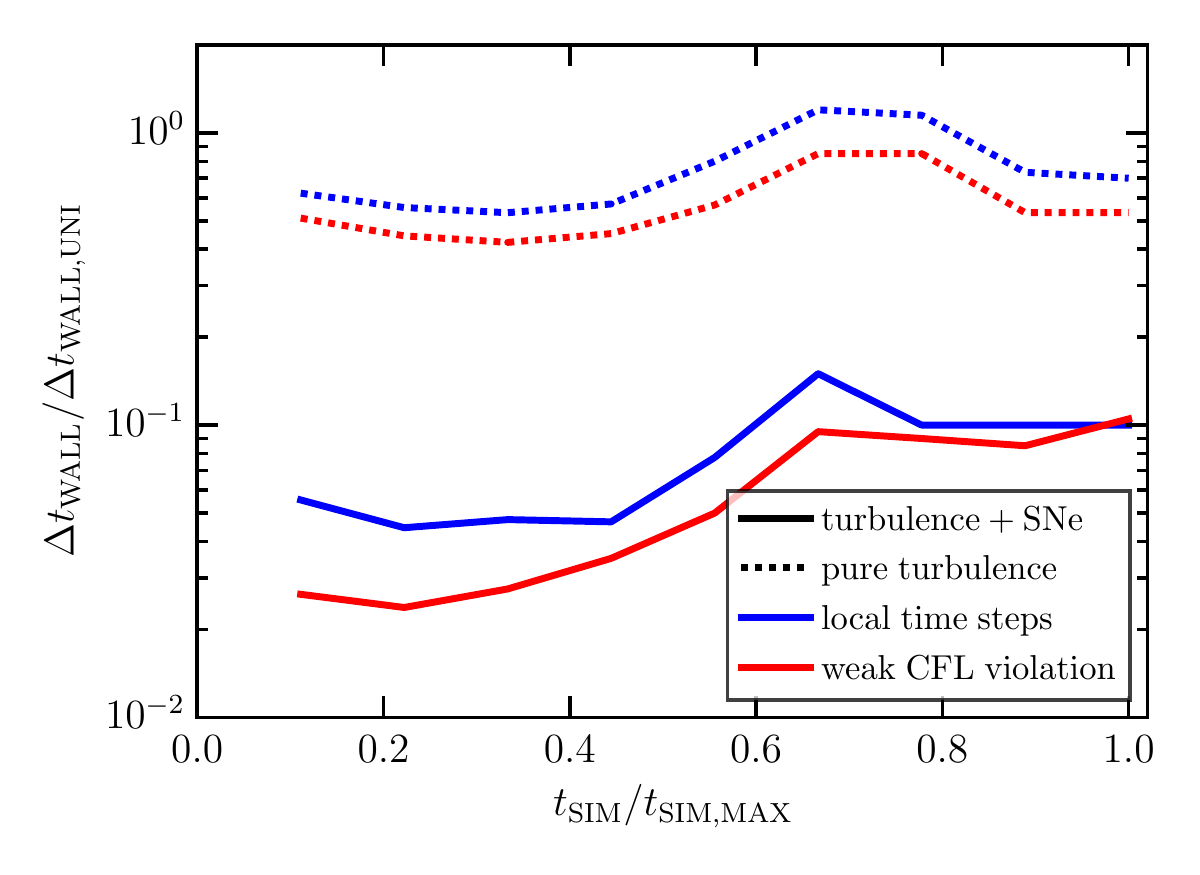}%
\caption{The ratio of the wall-clock times for tests with uniform time stepping and with local time stepping for decaying turbulence tests. Dotted lines show the original Kritsuk et al test \cite{knc11}, while solid lines show the same tests with additional point explosions (``supernovae'') allowed to go off in dense gas, as a more realistic interstellar medium model.}
\label{fig:tturb}
\end{figure}

Timing of the decayed turbulence tests is shown in figure \ref{fig:tturb}. Isothermal turbulence does not exhibit large variations in local time steps across the computational volume, and performance gains in simulating it with local time steps are moderate. Realistic interstellar medium, however, includes stars that explode as supernovae and inject energy and momentum in the surrounding gas. To mimic such an environment, we also allowed point explosions to occur in the decaying turbulence box in regions where gas becomes denser than the mean density in the volume by a factor of $\geq 100$. After each explosion the hot gas behaves adiabatically until it cools enough so that its temperature falls below 100 times the temperature in the isothermal gas; in other words, the equation of state in the test with explosions is $S={\rm const}$ for $T>100\;T_{\rm ISO}$ and $T=T_{\rm ISO}$ otherwise, where $T_{\rm ISO}$ is the temperature in the original Kritsuk et al isothermal test \cite{knc11}.

Since hot gas appearing in such explosions results in a wide distribution of sound speeds and thus local time step constraints, performance gain in this test is large. Again, just as in the point explosion test shown above, violating the CFL condition weakly does not produce significant gains in the performance. These tests illustrate that local time steps result in the performance gain only when the problem has a wide range of maximal time step constraints for individual cells.

\section{How Bad Are Weak CFL Violations?}
\label{sec:weak}

Given that strict enforcement of the CFL condition is a non-trivial task and requires additional communications and computational effort, one may be tempted to allow for weak CFL violations, i.e. strictly enforcing the local time step constraint (\ref{eq:lns}), but not the full CFL condition. In order to check whether allowing weak CFL violation leads to instabilities or numerical artifacts, we have performed a number of tests with weak CFL violations\footnote{I.e., with local time step constraint (\ref{eq:lns}) enforced with $C_{\rm CFL}=0.8$, but the CFL condition (\ref{eq:cfl}) not enforced.}. We find that the numerical solutions remain stable in all cases, but exhibit numerical artifacts of the kind discussed above for the case of point explosion (see fig.\ \ref{fig:wcsed}). Specifically, we found that the numerical solutions for the following test problems (in 1D, 2D, and 3D) remain stable  when the CFL condition is violated weakly: advection of a single wave, Kelvin-Helmholtz instability, the Sedov-Taylor point explosions, Riemann shock tube test. 
 
 In addition, we have also implemented a parabolic PDE solver to model nonlinear diffusion in our experimental code, and found that tests of diffusion calculations also remain stable under weak CFL violations. Thus, we conclude that solutions that violate CFL weakly are not necessarily unstable, although their numerical stability cannot be guaranteed and need to be tested for in each specific case. 

Examples in figures\ \ref{fig:wcsed} and \ref{fig:wcsed2} show that artifacts introduced by weak CFL violations can be arbitrary large. However, such errors decrease when variation of the locally allowed time step becomes smaller. One example of a realistic system with relatively weak variation of time steps is developed supersonic turbulence studied in the previous section. 

\begin{figure}[th]
\centering%
\includegraphics[width=\hsize]{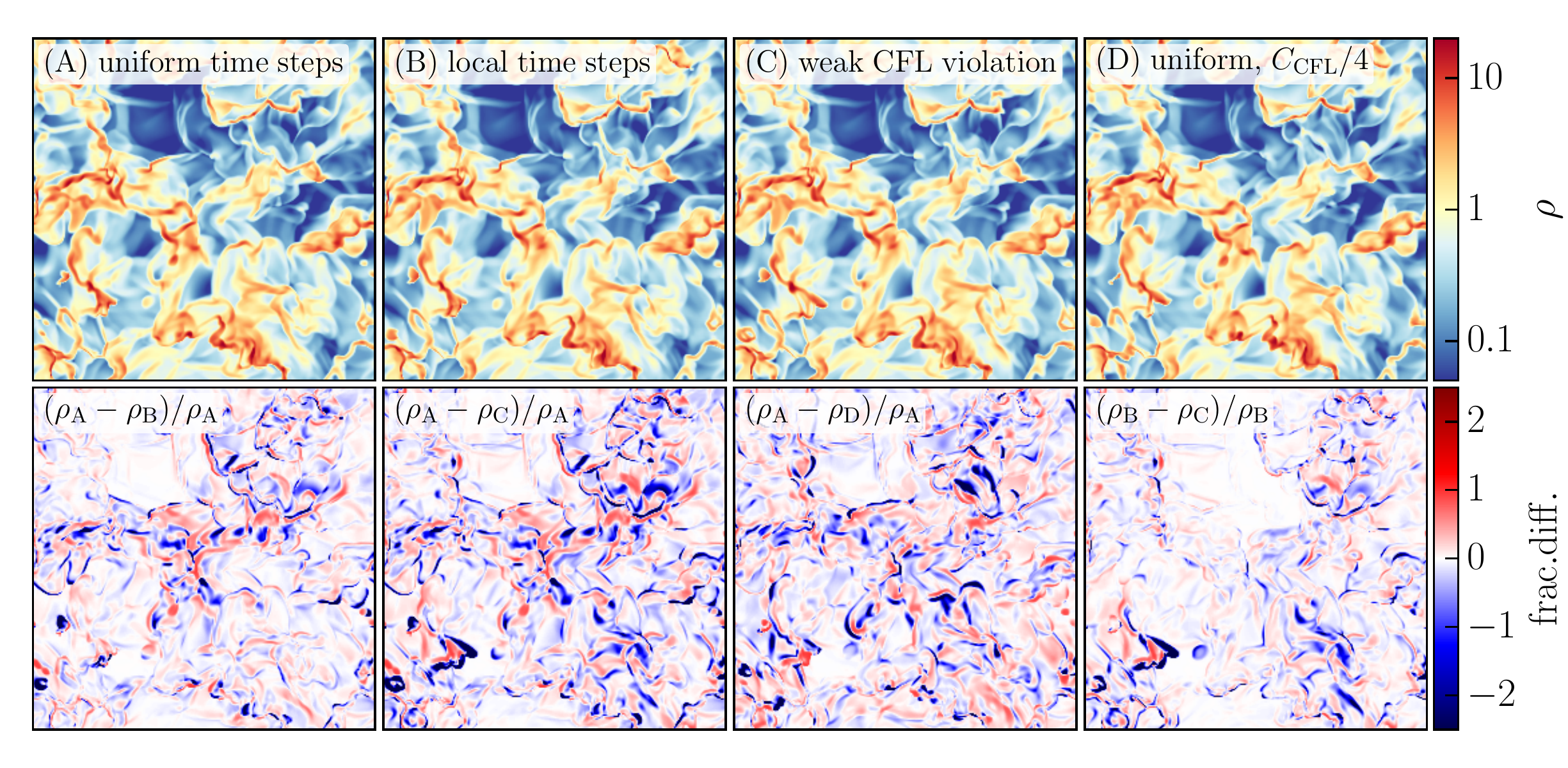}
\caption{Comparison of the decaying turbulence simulations with different time stepping and with and without strict enforcement of the CFL condition. The two simulations with uniform steps adopt CFL coefficients of $\cfl = 0.5$ (run ``A'') and $\cfl=0.125$ (run ``D''). The upper row of panels show density slices in different runs, as indicated in the upper left corner of each panel. The bottom row of panels shows the maps of fractional differences between the runs as indicated in the upper left corner of each panel. In all cases the differences are comparable.}
\label{fig:wscfl}
\end{figure}

In figure \ref{fig:wscfl} we illustrate the errors introduced by weak CFL violations by comparing four decaying turbulence simulations: (A) run with globally uniform time stepping and strict enforcement of the CFL condition, (B) run with local time step constraint only and strict enforcement of the CFL condition, (C) run with local time stepping which allows for weak CFL violations, and (D) run with the globally uniform time steps that are further reduced by a factor of 4 over what is required by the CFL condition. 

\begin{figure}[ht]
\centering%
\includegraphics[width=0.5\hsize]{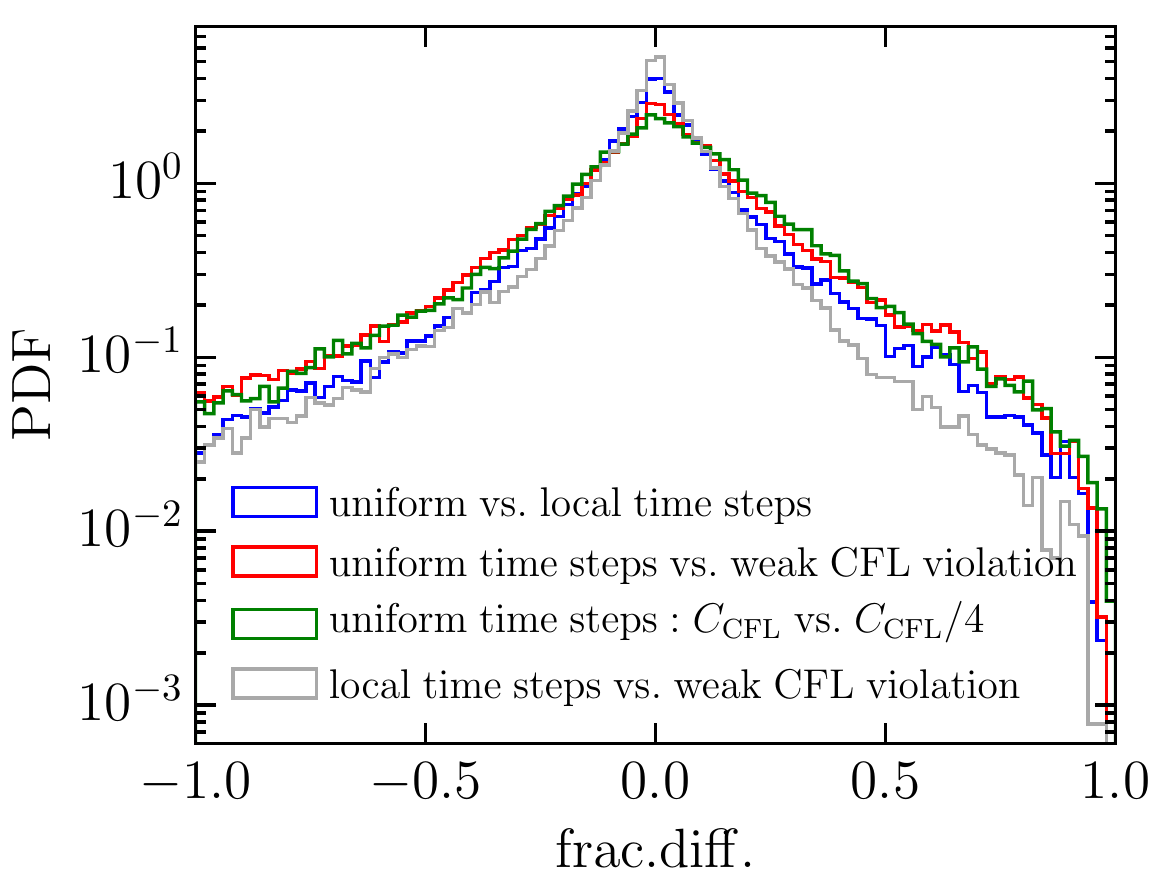}%
\includegraphics[width=0.5\hsize]{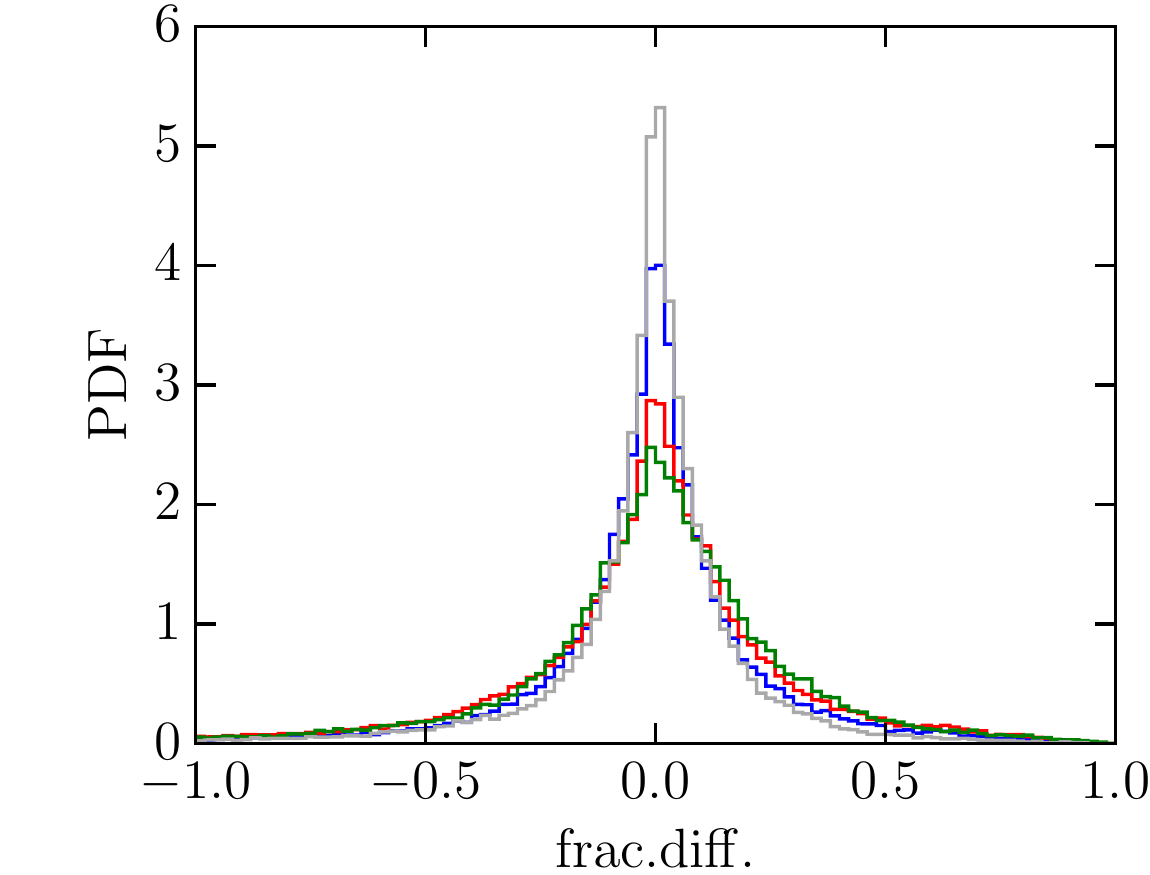}%
\caption{Distribution of pairwise fractional differences between the four tests from fig.\ \ref{fig:wscfl} with logarithmic (left) and linear (right) vertical scale. Confirming the visual impression, distributions of differences for all runs are similar.}
\label{fig:tdif}
\end{figure}

The figure shows that the difference between the LTS run with weak CFL violation and run with uniform steps or LTS with strict CFL enforcement are comparable to the differences between uniform time step runs with different CFL coefficients. The supersonic turbulent flows in these simulations result in wide variation of gas densities and complicated network of shocks and hierarchical density structures. Visually, all large-scale structures are similar in all four runs, but fractional differences on cell scales arise because these structures are slightly misplaced relative to each other. The distributions of fractional differences shown in the bottom row of figure \ref{fig:wscfl} are compared in figure \ref{fig:tdif}. In the language of waves, all the waves are similar in all four cases, but their phases are slightly different in four solutions. Apparently, the artifacts we demonstrated in fig. \ref{fig:wcsed} do not contribute substantially to the overall error in this solution, but lead to small differences in the location of waves.

\begin{figure}[ht]
\centering%
\includegraphics[width=0.5\hsize]{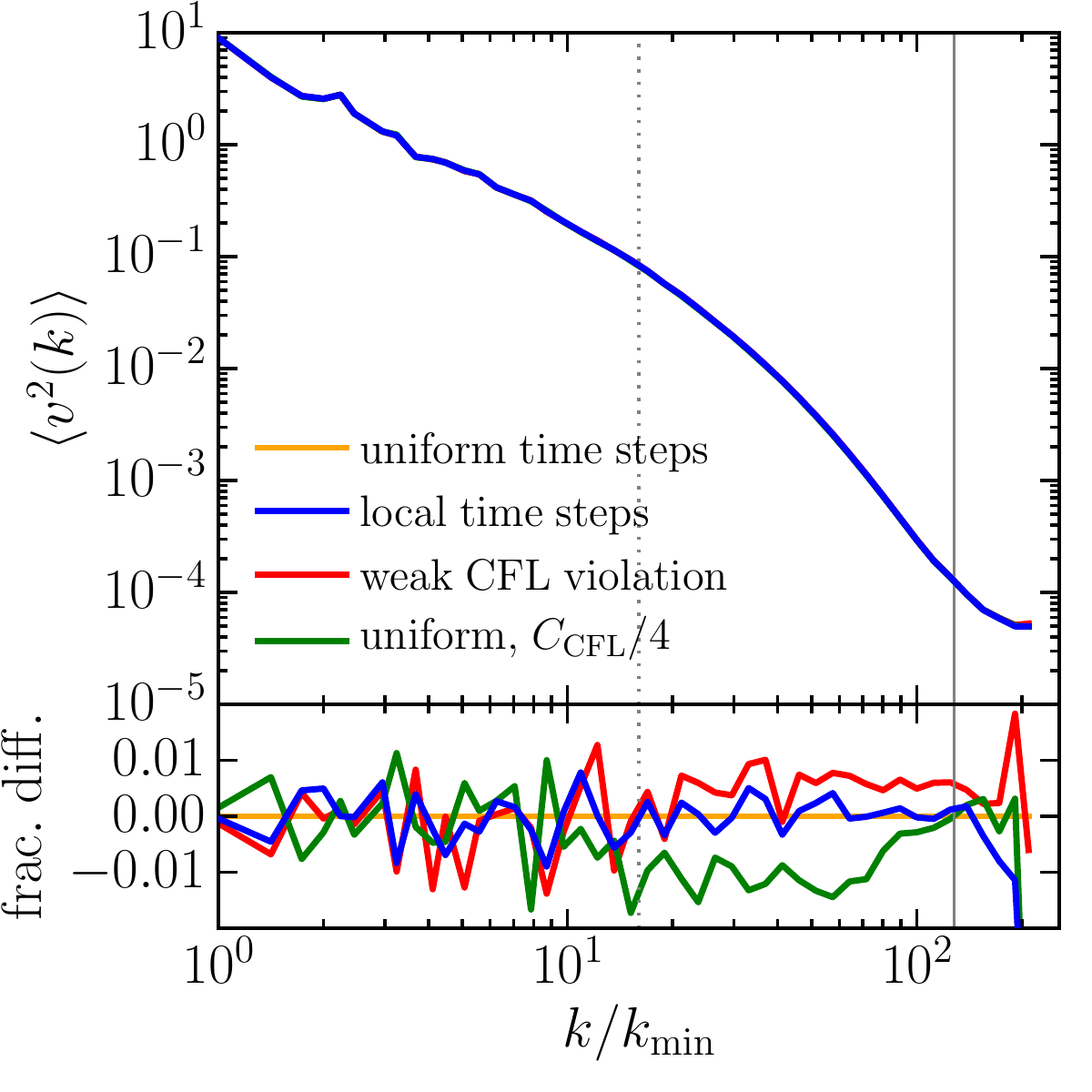}%
\includegraphics[width=0.5\hsize]{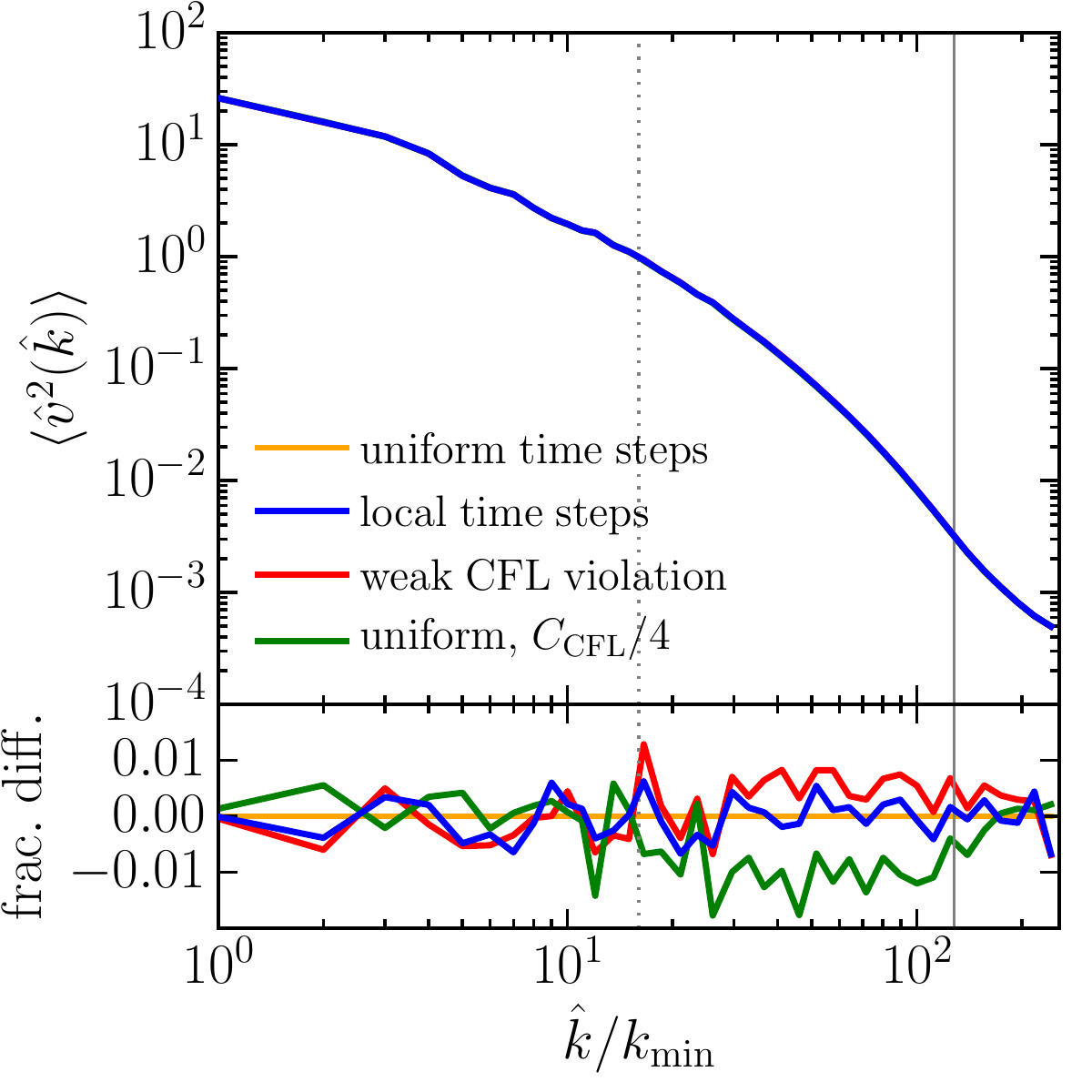}%
\caption{Power spectra of the density field in the four isothermal turbulence tests (top panels)  from figure\ \ref{fig:wscfl} and their fractional differences (bottom panels) with the case of uniform time stepping. Vertical solid and dotted lines show the Nyquist frequency and the frequency corresponding to the patch size (16 cells). Left panel shows the standard velocity power spectrum $P(k=|\vec{k}|)=\langle v^2(k) \rangle$ with modes averaged in spherical $[k,k+dk]$ shells in the Fourier space, while the right panel uses the ``Manhattan distance'' in $k$-space $\hat{k}=|k_x|+|k_y|+|k_z|$ to define the wavenumber of each mode.}
\label{fig:ps}
\end{figure}

For simulations with chaotic nonlinear processes, such as the decaying turbulence test, one is often most interested in statistical properties of the solution rather than exact location of waves. In particular, power spectrum is a statistic commonly used to quantify turbulent flows. 
 Given that in our experimental code time steps change abruptly across neighboring cubic patches, numerical artifacts due to local time steps could conceivably result in features in the power spectrum on the patch scale. Power spectra for the four test runs from figure\ \ref{fig:wscfl} are shown in figure \ref{fig:ps}. The figure shows both the standard spherical-in-$k$-space velocity power spectrum $P(k=|\vec{k}|)=\langle v^2(k) \rangle$, and the power spectrum $P(\hat{k})$ that uses ``Manhattan distance'' in $k$-space, $\hat{k}=|k_x|+|k_y|+|k_z|$.

The figure shows that there are no significant artifacts introduced at the patch scale in any of the test runs. Here again the differences between simulations with the globally uniform time step and LTS runs are similar to those between runs with uniform time step but different CFL coefficients. In particular, power spectra for both runs with weak CFL violations and with uniform time steps with $\cfl=0.125$ case start deviating systematically by $\sim$1\% from the $\cfl=0.5$ uniform time step simulation on scales smaller than the patch scale.

Thus, weak CFL violations do not necessarily result in catastrophic errors in the decaying turbulence solution. However, this does not mean that it will generally be true in all cases. If one decides to allow for weak CFL violations, thorough testing of the effects of such choice will be required to prove fidelity of the solution. This is because the CFL condition is a necessary condition for numerical \emph{stability and convergence}. A solution weakly violating the CFL condition is not guaranteed to converge to the correct one, as figs.\ \ref{fig:wcsed} and \ref{fig:wcsed2} demonstrate.

Given that thorough testing to ensure correctness of the solution may not be feasible, generally simulations should enforce the CFL condition strictly. The algorithm presented in this paper illustrates how this can be done in a parallel implementation of the LTS approach.

\section{Conclusions}

Just like the adaptive mesh refinement allows one to concentrate computational resources where they are needed, local time stepping allows to adjust allocation of resources to the actual variation in characteristic time scales exhibited by a numerical solution as it evolves. When time steps of individual resolution elements or of small groups of resolution elements (``patches'' in our specific implementation) vary locally, the strict enforcement of the Courant-Friedrichs-Lewy condition can be achieved by forcing regions to never advance pas a neighbor that steps with its own, CFL-limited time step (condition \ref{eq:cfl}). This condition is  different from the ``update criterion'' (or ``evolve condition'') discussed in previous studies \cite{lgm07,dkt07,lgm08,d14,cdm15}. However, enforcing the CFL condition exactly in a conservative numerical scheme with local time steps, can be non-trivial. 

The local time step constraint (\ref{eq:lns}) is a necessary stability condition for any numerical scheme. Schemes that enforce it but violate the CFL condition (\ref{eq:cfl}) do not ensure the correct propagation speed for all waves, the issue we call weak CFL violation, and thus result in artifacts at the cell interfaces at which time steps differ significantly. Such schemes can be stable, or, at least, are not necessarily violently unstable. We find that in a large set of common fluid dynamics tests and in tests with a nonlinear diffusion solver, solutions that allow weak CFL violations remain numerically stable. However, unless weak CFL violations are shown explicitly to be tolerable, simulations should in general enforce the CFL condition strictly. 

We also find that in all our tests, when weak CFL violations do not introduce large numerical artifacts, they also do not result in any significant gain in execution time. We cannot prove that this finding applies generally, but the tests we considered serve as examples that violating the CFL condition does not necessarily decrease the execution time.

While test results we presented in this paper were obtained with the hyperbolic solver for fluid dynamics problems, we expect our conclusions to be applicable to other PDEs as well. In particular, we have also implemented a nonlinear diffusion (parabolic) solver in our experimental code, and reproduced all of the results and issues discussed in this paper with it. Hence, we expect our conclusions to be sufficiently general for most explicit conservation laws.

{\bf Acknowledgements}
Ideas that prompted the work described in this paper were born during several discussions with Troels Haugb{\o}lle, {\AA}ke Nordlund, Michael Norman, and Romain Teyssier on the future exa-scale high performance computing. We also thank Troels Haugb{\o}lle, Michael Norman, and Jim Stone for comments on the early draft and Alexei Kritsuk for providing access to the initial conditions of isothermal turbulent test used in the comparison project of \cite{knc11}. Finally, we acknowledge critical, but constructive comments from two anonymous referees that allowed us to substantially improve the original manuscript.

Fermilab is operated by Fermi Research Alliance, LLC, under Contract No. DE-AC02-07CH11359 with the United States Department of Energy. This work was supported in part by a NASA ATP grant NNH12ZDA001N, NSF grant AST-1412107, and by the Kavli Institute for Cosmological Physics at the University of Chicago through grant PHY-1125897 and an endowment from the Kavli Foundation and its founder Fred Kavli. The simulation presented in this paper have been carried out using the Midway cluster at the University of Chicago Research Computing Center, which we acknowledge for support.

\appendix

\section{Gas Dynamic Solver}
\label{sec:axcode}

The gas dynamics solver used in our experimental code is adopted from the publicly available cosmological code RAMSES \cite{t02}. The solver solves Euler equations in the conservative form,
\[
  \frac{\partial}{\partial t}\begin{bmatrix}\rho\\ \rho\vec{u}\\ e\end{bmatrix} + \frac{\partial }{\partial \vec{x}}\begin{bmatrix}\rho\vec{u}\\ \rho\vec{u}\otimes\vec{u}\\ (e+p)\vec{u}\end{bmatrix} = 0,
\]
where $\rho$, $\vec{u}$, $e$, and $p$ are the gas density, velocity, energy density, and pressure respectively, in 1D, 2D, or 3D on a spatially uniform grid. It implements a second order accurate in time and space Godunov-type directionally unsplit finite volume scheme that uses the HLLC (Harten-Lax-van Leer-Contact) Riemann solver \cite{tss94}. The reconstruction on the faces is done using piece-wise linear interpolation (PLM) with the "minmod" limiter to insure the monotonicity of the solution.

\section{Execution Model}
\label{sec:app}

It is instructive to consider explicitly the achieved gains and incurred costs associated with local time stepping and the enforcement of the CFL condition. Let's consider a computational volume divided into a number of domains with (on average) $N_V$ patches per domain, each patch taking wall-clock time $T_W$ to solve the gas dynamics on. Domains communicate by sending messages to each other (for example, via MPI, although the specific communication protocol is not important here); these messages are exchanged between $N_B$ boundary patches that have one or more of their boundaries aligned with domain boundaries (for well formed domains, one can expect $N_B \propto N_V^{2/3}$ in 3D), so that boundary patches must use communication to exchange information with their neighbors (as opposite to internal, non-boundary patches that can directly access their neighbor data at a negligible cost). Because solving gas dynamics equations on each patch requires boundary data from neighbors, each boundary patch also takes, on average, wall-clock time $T_C$ to communicate with its neighbors. 

Hence, a single globally uniform time step takes a wall-clock time $N_VT_W + N_B T_C$ to complete. Let's also assume that the whole simulation takes $2^m$ globally uniform time steps, so that the total wall-clock time taken by a conventional simulation with uniform time steps is
\begin{equation}
  T_{\rm UTS} = 2^m\left(N_VT_W + N_B T_C\right).
  \label{eq:tuts}
\end{equation}

When the local time stepping is allowed, individual patches take time steps of duration $2^j$, where $j$ runs from 0 (the smallest allowed time step, equal to the time step of a simulation with globally uniform time steps) to $m$ (some patches may finish the whole simulation in just one time-step). The fraction of patches that evolve with the time step $2^j$ is $f_j$, so that
\[
  \sum_{j=0}^m f_j = 1.
\]
(the globally uniform time-step case corresponds to $f_0=1$ and $f_j=0$ for all $j>0$). In our approach, in the worst case scenario each patch stepping with time-step with index $j$ may send "pull/drop" messages to all patches that evolve with a larger time-step (a "pull" message goes to immediate neighbors, but then the follow-up "drop" messages go to all other patches with larger time-steps), so the number of additional messages at step index $j$ is $f_j \sum_{i=j+1}^m f_i$. These additional messages are much smaller than messages that update boundary data for patches, but, again in the worst case scenario of a fast network and a low quality MPI implementation, the MPI overhead of message handling may be larger than the actual time for sending the data over the network, so that a small "pull/drop" message can, in principle (although, this really should be quite unlikely), take as much wall-clock time as the data exchange message. Hence, in the LTS case, the total wall-clock time becomes
\begin{equation}
  T_{\rm LTS} \leq N_VT_W\sum_{j=0}^m 2^{m-j} f_j + N_B T_C \sum_{j=0}^m 2^{m-j} f_j \left(1 + \sum_{i=j+1}^m f_i\right).
  \label{eq:tlts}
\end{equation}

A natural scaling limit for any simulation is the number of cores at which the work time per patch is equal to the communication time, i.e. $N_VT_W = N_B T_C$ (we assume that $T_C \gg T_W$, otherwise the parallel scaling of the code is unlimited, since by definition $N_B \leq N_V$). In that case the wall-clock time taken by the globally uniform time step simulation is
\[
  T_{\rm UTS} = 2^{m+1} N_VT_W 
\]
and the time taken by the LTS simulation is
\[
  T_{\rm LTS} \leq N_VT_W \sum_{j=0}^m 2^{m-j} f_j \left(2 + \sum_{i=j+1}^m f_i\right).
\]
Obviously, the latter depends on the distribution of patches over their time steps $f_j$. The ratio of the two can be easily computed in this limit,
\begin{equation}
  R_{\rm LTS} \equiv \frac{T_{\rm LTS}}{T_{\rm UTS}} \leq \sum_{j=0}^m 2^{-j} f_j \left(1 + \frac{1}{2}\sum_{i=j+1}^m f_i\right).
  \label{eq:rlts}
\end{equation}
One can observe that
\[
  R_{\rm LTS} \leq f_0 + \frac{1}{2}\sum_{j=1}^m 2^{1-j} f_j + \frac{1}{2}f_0(1-f_0) + \frac{1}{2}\sum_{j=1}^m 2^{1-j} f_j \sum_{i=j+1}^m f_i,
\]
and $\sum_{j=1}^m 2^{1-j} f_j \leq \sum_{j=1}^m f_j = 1 - f_0$, while $\sum_{i=j+1}^m f_i = 1 - \sum_{i=0}^j f_i \leq 1-f_0$, so that for any distribution $f_j$,
\[
  R_{\rm LTS} \leq f_0 + \frac{1}{2}(1-f_0) + \frac{1}{2}f_0(1-f_0) + \frac{1}{2}(1-f_0)^2 = 1.
\]
In other words, the LTS simulation never takes longer to complete than a globally uniform time-step case (and the equality is only achieved when $f_0=1$ ans $f_j=0$ for all $j>0$, i.e.\ when all patches have exactly the same time-steps). Notice, that this conclusion is actually independent of our assumption made above that time-steps are discretized as powers of two, and will remain valid for any distribution of time steps $\Delta t_j$ such that $\Delta t_j \leq \Delta t_{j+1}$.

Just to illustrate the case where the gains in using LTS are significant, let us assume that patches are distributed approximately uniformly over different time steps, i.e.\ $f_j = 1/(m+1)$. In that case 
\[
  R_{\rm LTS} \approx \frac{1}{m+1}\sum_{j=0}^m 2^{-j} \left(1 + \frac{m-j}{2(m+1)} \right) = \frac{(3m+1-m 2^{-m})}{(m+1)^2} \approx \frac{3}{m}
\]
if $m \gg 1$. Without "pull/drop" messages (i.e., with allowing weak CFL violations), the second term in the parenthesis inside the sum would disappear, and $R_{\rm LTS} \approx 2/m$ for large $m$. I.e., weak CFL violations save about 30\% of the wall-clock time, which is broadly consistent with Fig.\ \ref{fig:tsed} and \ref{fig:tturb} (recall that the exact value of 30\% is obtained for the specific case of $f_j=1/(m+1)$ and $N_VT_W = N_B T_C$).

The model presented above is valid for non-conservative schemes, in which only boundary data are exchanged between neighboring patches. For a conservation law, flux also need to be exchanged between neighboring patches, because, in general, if two patches step with different time steps, fluxes computed on the two sides of one interface will not be the same, and in a conservative scheme one of them needs to be discarded, while the other one needs to be transferred to the neighbor (over the network in case the two neighboring patches are located on two separate domains). 

In a general case it is difficult to incorporate this cost in the model above. However, in the limit when the MPI message handling overhead dominates over the actual data transfer, one can obtain an upper limit on the additional cost of flux exchange by noticing that the flux between two neighboring patches must be exchanged every time a "pull" message is sent: Patch0 in the right panel of Fig.\ \ref{fig:sc2} must send its flux to Patch 1 for the latter to be able to step. Hence, the cost of flux exchange (again, only in the limit of MPI overhead dominating) cannot exceed the cost of "pull/drop" messages, i.e., comparing with equation (\ref{eq:rlts}),
\[
  R_{\rm LTS}^{+{\rm FLUX}} \leq \sum_{j=0}^m 2^{-j} f_j \left(1 + \sum_{i=j+1}^m f_i\right),
\]
or, in a more general case of $N_VT_W \neq N_B T_C$,
\begin{equation}
  R_{\rm LTS}^{+{\rm FLUX}} \leq \sum_{j=0}^m 2^{-j} f_j \left(1 + \frac{N_B T_C}{N_VT_W}\sum_{i=j+1}^m f_i\right).
  \label{eq:rltsflux}
\end{equation}


\def\aap{A\&A}
\def\apj{ApJ}
\def\apjl{ApJL}
\def\apjs{ApJS}
\def\mnras{MNRAS}
\def\na{New Astronomy}

\bibliographystyle{model1-num-names}
\bibliography{paper.bib}







\end{document}